\documentclass[apj]{emulateapj}

\DeclareRobustCommand{\rchi}{{\mathpalette\irchi\relax}}
\newcommand{\irchi}[2]{\raisebox{\depth}{$#1\chi$}} 

\usepackage[colorlinks=false]{hyperref} 
\usepackage{tabularx}
\usepackage{threeparttable}
\usepackage{multirow}

\usepackage{amssymb}
\usepackage{amsmath}

\usepackage{graphicx,color}
\usepackage{graphicx,subfigure}

\shorttitle{Performance and accuracy of \textsc{relxill}}
\shortauthors{Choudhury et al.}

\begin{document}
	
	
	\title{Testing the performance and accuracy of the \textsc{relxill} model for the relativistic X-ray reflection from accretion disks}

	
	\author{Kishalay Choudhury\altaffilmark{1}}
	
	\author{Javier A. Garc{\'{\i}}a\altaffilmark{2,3,4,$\dagger$}}
	
	\author{James F. Steiner\altaffilmark{5,$\ddagger$}}
	
	\author{Cosimo Bambi\altaffilmark{1,6,$\dagger$}}
	
	\altaffiltext{1}{Center for Field Theory and Particle Physics and Department of Physics, Fudan University, 200433 Shanghai, China}
	
	\altaffiltext{2}{Cahill Center for Astronomy and Astrophysics, California Institute of Technology, Pasadena, CA 91125, USA}
	\altaffiltext{3}{Harvard-Smithsonian Center for Astrophysics, 60 Garden St., Cambridge, MA 02138 USA}
	\altaffiltext{4}{Remeis Observatory \& ECAP, Universit{\"a}t Erlangen-N{\"u}rnberg, Sternwartstr.~7, 96049 Bamberg, Germany}
	
	\altaffiltext{5}{MIT Kavli Institute for Astrophysics and Space Research, MIT, 70 Vassar Street, Cambridge, MA 02139}
	
	\altaffiltext{6}{Theoretical Astrophysics, Eberhard-Karls Universit{\"a}t T{\"u}bingen, 72076 T{\"u}bingen, Germany}
	
	\altaffiltext{$\dagger$}{Alexander von Humboldt Fellow}
	\altaffiltext{$\ddagger$}{Einstein Fellow}

	\begin{abstract} 
		
		The reflection spectroscopic model {\sc relxill} is commonly implemented in studying relativistic X-ray reflection from accretion disks around black holes. We present a systematic study of the model's capability to constrain the dimensionless spin and ionization parameters from $\sim$6,000 \textit{NuSTAR} simulations of a bright X-ray source employing the lamppost geometry. We employ high count spectra to show the limitations in the model without being confused with limitations in signal-to-noise. We find that both parameters are well-recovered at 90\% confidence with improving constraints at higher reflection fraction, high spin, and low source height. We test spectra across a broad range-- first at 10$^6-$10$^7$ and then $\sim$10$^5$ total source counts across the effective 3--79~keV band of {\it NuSTAR}, and discover a strong dependence of the results on how fits are performed around the starting parameters, owing to the complexity of the model itself. A blind fit chosen over an approach that carries some estimates of the actual parameter values can lead to significantly worse recovery of model parameters. We further stress on the importance to span the space of nonlinear-behaving parameters like $log~\xi$ carefully and thoroughly for the model to avoid misleading results. In light of selecting fitting procedures, we recall the necessity to pay attention to the choice of data binning and fit statistics used to test the goodness of fit by demonstrating the effect on the photon index $\Gamma$. We re-emphasize and implore the need to account for the detector resolution while binning X-ray data and using Poisson fit statistics instead while analyzing Poissonian data.
	
	\end{abstract} 
	
	
	\maketitle

	\renewcommand{\thefootnote}{\roman{footnote}}

	\section{Introduction}\label{intro}
	
	\setcounter{footnote}{0} 
	The modeling of the X-ray reflection spectrum is a very important method for
	understanding the physics of accreting compact objects.  X-ray spectra from
	active galactic nuclei (AGN) and X-ray binaries often show evidence of
	interaction between radiation emitted near the compact object and the nearby
	gas, which leads to signatures imprinted on the observed spectrum. Modeling the
	observed X-ray reflection features can lead to important constraints on the
	ionization state of the inner accretion disk
	\citep{Ross1999,GK2010,GKM2011,Garcia2013}. The reflecting region of the disk
	may be subject to relativistic effects that can blur and distort the emission
	features \citep{Fabian1989,Laor1991}, leading to measurements of inner disk
	radii and black hole spin \citep{BR2006,Rey2008}.  The most notable feature is
	the Fe-K$\alpha$ emission complex (e.g., \citealt{LW1988,GR1988,Fabian89,GF1991}).  
	An
	astrophysical black hole in general relativity is completely specified by its
	angular momentum $J$ and its mass $M$ \citep{Kerr1963}\footnote{Because of the difference between the masses of electrons and protons, astrophysical objects can have a non-vanishing equilibrium electric charge. However, for macroscopic objects, the value of the equilibrium electric charge is small, and completely negligible in the background metric.}. 
	Black hole
	spin, defined by the dimensionless spin parameter $a_* = cJ/GM^2$ with
	theoretical values $|a_*| \leqslant 1.0$, is arguably the most important
	parameter whose estimate is affected by strong field gravity near the black
	hole. 
	One of the best-known AGN systems and the first to have observationally-confirmed broad iron line detection is
	the Seyfert I galaxy MCG-6-30-15 \citep{Tanaka1995,Iwa1999}, for which the Fe
	emission appears to be broad and skewed well beyond the instrumental
	resolution.  Such an X-ray reflection spectrum has been observed from accretion
	disks around numerous black holes \citep[e.g., see][]{Rey2013}.  

	Numerous reflection model computations have been published over the past two decades, with 
	the most notable including \texttt{PEXRAV} \citep{Mag1995}, \texttt{REFLIONX}
	\citep{RF2005}, and \texttt{XILLVER} \citep{GK2010,Garcia2013}. These models
	were originally decoupled from the relativistic smearing associated with strong gravity and implemented in broadening kernels such as
	\texttt{DISKLINE} \citep{Fabian1989}, \texttt{LAOR} \citep[][extended to
	\texttt{KDBLUR} later]{Laor1991}, \texttt{KERRDISK} \citep{BR2006}, \texttt{KY}
	\citep{Dovciak2004}, and \texttt{RELLINE} \citep{Dauser2010,Dauser2013}.  These have been applied in a great many observational papers, (e.g., \citealt{Miller2008,Steiner2011,Rey2012,Fabian2012b,Dauser2012}). 
	
	The X-ray blurring code \textsc{relxill}
	presented by \cite{Garcia2014} is the current most advanced relativistic reflection
	model which has addressed many of the deficiencies of previous models. It is the
	result of the angle-dependent reflection code \texttt{XILLVER} \citep{Garcia2013} convoluted with
	the relativistic blurring code \texttt{RELLINE} \citep{Dauser2013}. \texttt{XILLVER} uses the
	atomic data of \texttt{XSTAR} \citep{KB2001} to calculate the specific
	intensity of the radiation field as a function of energy, position in the
	accretion disk, and emission angle. Lamp-post geometry
	\citep{Matt1991,Mar1996,Frank2002,Dauser2013} which describes an isotropically irradiating
	primary X-ray source located on the rotation axis of the black
	hole has been implemented as \texttt{relxilllp} in the 
	\textsc{relxill} model family.  Given the progress made on reflection computations, and the complexity of the 
	these calculations, it is important to test and understand how
	reliably, in an idealized case, the model performs in yielding estimates of spin and other quantities of interest.

	The \textit{Nuclear Spectroscopic Telescope Array} (\textit{NuSTAR}; \citealt{nustar2013}) mission is
	currently the best resource for the study of relativistic reflection. Its broad bandpass enables
	simultaneous measurements of both the iron emission and the Compton hump.
	Examples of recent, high-count reflection studies using 
	\textit{NuSTAR} data are spin measurements for the high-mass X-ray binary (HMXB)
	Cygnus~X-1 \citep{Walton2016} in the soft state, the low-mass X-ray
	binary (LMXB) GS~1354-645 \citep{Batal2016} in the hard state, and for the LMXB GX~339-4 in the very high state along with {\it Swift} XRT data \citep{Parker2016}. All works show
	that good constraints on spin can be achieved using \textit{NuSTAR}
	for such bright sources, and their results agree well with previous
	observations. A more recent work with V404~Cygni has established the source's first ever spin constraint using {\it NuSTAR} \citep{Walton2017b}. Hard X-ray observations of accreting black holes provide a probe
	of the inner regions of the accretion disk where strong gravity prevails
	\citep{RM2006,Bambi2016}.

   Using data with good signal and enough counts is a very common practice for simulations in order to test a model's efficiency. But one also needs to be cautious of systematic errors that may be introduced as a consequence of the ubiquitous customs. \cite{cstat2009} presented a thorough analysis and stated via Monte Carlo tests
that performing fits employing $\rchi^2$-statistics at high counts can lead to heavy bias in contrast to the popular belief that the effect is minimal if a minimum-counts-per-bin sampling approach is taken up at high counts that falls in-line with expectancy from Cash statistics (\cite{Cash1979}). The latter statistic should be employed for analyzing Poisson-distributed data and has been shown in their work to yield unbiased parameter estimates at high counts. As explained in the paper, the ``approximated'' $\rchi_d{^2}$ test (followed as
the \textit{de facto} standard $\rchi^2$-test on \textsc{xspec})
can prove to underestimate values largely as counts in the spectrum increase,
and unless the number of data bins in the dataset are far less than $\sqrt{N}$,
where $N$ is the number of counts in the energy range considered in the fit,
the bias with high-count data will not cease to exist irrespective of sampling with high counts per bin in a
$\rchi^2$-test. In fact, the bias can be of the order of or even higher than
the statistical error, and has been shown to increase as the number of counts
increase. We will, however, show in the next section that the kind of binning we adopt avoids this problem for the total counts we work with. \label{humphrey}
\par

	
	In this paper, we aim to demonstrate how well \textsc{relxill} can constrain 
	spin and related spectral parameters under optimal conditions. The work is presented as follows: Section \ref{degen} describes the model, 
	software, and our methodology.  Our results are given in Section~\ref{disc}.  In Section~\ref{compare}, we also present a comparison of results from our fitting approach with that of a recent
	paper, and in Section \ref{StatTest} we further elucidate the role of proper data analysis techniques which are important when assessing subtle features in the reflection signal to infer physical constraints.   We present our conclusions in Section
	\ref{conc}.

	\section{Simulations as Proxy for Ideal Observational Data}\label{degen}

	Lamp-post geometry has been employed for explaining the observed X-ray
	spectra of numerous sources in the past \citep[e.g., ][]{Duro2011,WF2011,Dauser2012,Marin2012,Miller2015}, and also in a few recent analyses \citep{Beuchert2017,Walton2017a,Walton2017b}. We use \texttt{v0.4a}\footnote{Newer versions
		of \textsc{relxill} were made available during the course of this project. But
		our results are still applicable because of our source constraints and
		methodology adopted.} of the lamp-post model of \textsc{relxill} in this paper
	to carry out our analysis.  Our main goal is
	to determine the robustness of the model, when employing ideally for the simplistic case of
	an axisymmetric, stationary source irradiating isotropically.

	\begin{table*}
		\centering
		\caption{Parameter values used in \texttt{relxilllp} in combination with 9 input
			$a_*$ values. Inner and outer disk radii, source redshift and high-energy cutoff were kept at default values. Except source height, the
			remaining were used as fit
			parameters.}\label{params}
		
		\begin{threeparttable}   
			\begin{tabularx}{6.5in}{cccccc} \hline \\[0.2mm]
				
				\textbf{Parameter} & \textbf{Input Values} & \textbf{Fit Parameter} & \textbf{Units} \\[1mm] \hline \hline
				
				Source height ($h$) & 3 , 5 & No & $R_g \left(=\frac{GM}{c^2} \right) $ \\[0.2mm]  
				Inclination ($i$) & 15 , 45 , 75 & Yes & deg \\[0.2mm]
				Inner disk radius ($R_{\rm in}$) & 1 & No & $R_{\rm ISCO}$ \\[0.2mm]
				Outer disk radius ($R_{\rm out}$) & 400 & No & $R_g$ \\[0.2mm]
				Redshift ($z$) & 0.0 & No & -- \\[0.2mm]
				Photon index ($\Gamma$) & 1.4 , 2.0 , 2.6 & Yes & -- \\[0.2mm]
				Log ionization ($log~\xi$) & 1.0 , 2.3 , 3.6 & Yes & -- \\[0.2mm]
				Iron abundance ($A_{\rm Fe}$) & 1 , 5 , 10 & Yes & solar \\[0.2mm]
				High-energy cutoff ($E_{\rm cut}$) & 300 & No & keV \\[0.2mm]
				Reflection fraction ($R_f$) & 0.5 , 1.0 , 5.0 , 10.0 & Yes\footnote{We checked the estimates on $R_f$ based on the dispersion in best-fit data. Not only are the errorbars on $R_f$ low, but also there is no significant difference in recovering all $R_f$ values between both $h$. The 90\% confidence is very narrow, and thus, $R_f$ can be assumed to be a non-contributing fit parameter in terms of underlying physics rather than the statistics involved.} & -- \\[1mm] \hline
				
			\end{tabularx}
			\vspace*{2mm}   
		\end{threeparttable}
	\end{table*}

	For this paper, we have simulated observational data using the {\tt fakeit}
	routine in \textsc{xspec} \citep{xspec} \texttt{v12.9.0i}, implemented with the
	Python interface \texttt{PyXspec} \citep{pyxspec} \texttt{v1.1.0}. The
	simulations use the instrumental response of the FPMA of \textit{NuSTAR} in the
	3--79~keV energy band.  Ancillary response and background files were
	selected assuming a circular extraction region with 60$''$ radius centered 60$''$
	off-axis. All the instrumental files used here are available on the official
	website of \textit{NuSTAR}\footnote{Point source simulation files,
		\url{http://www.nustar.caltech.edu/page/response_files}}.
	
	The resolution of the \textit{NuSTAR} detectors is 0.4/0.9~keV at
	6/60~keV. 
	Based on this, we approximate the detector resolution as scaling as $E^{2/5}$ with a constant offset, and use the FTOOLS subroutine \texttt{GRPPHA} to bin 
	our spectra so that the energy resolution is oversampled by a factor of 3. Approximately 90\% of the grouped bins had $>50$~cts~bin$^{-1}$, with the typical number of counts per bin even higher in harder input spectra compared to softer ones. Readers may refer to \cite{Kaastra2016} for a more operationally crafty optimal binning method laid down for X-ray data analysis. This concerns both signal-to-noise and binning according to detector resolution. It adopts a variable binning scheme, thereby greatly reducing response matrix size and the number of model bins to save a lot of storage and computational time. This is, however, a much more sophisticated method than the one we have employed here. Nevertheless, we find that our approach serves well the main goals of the present paper.
	
	\begin{figure*}
		\centering
		
		\subfigure{
			\includegraphics[width=0.475\textwidth]{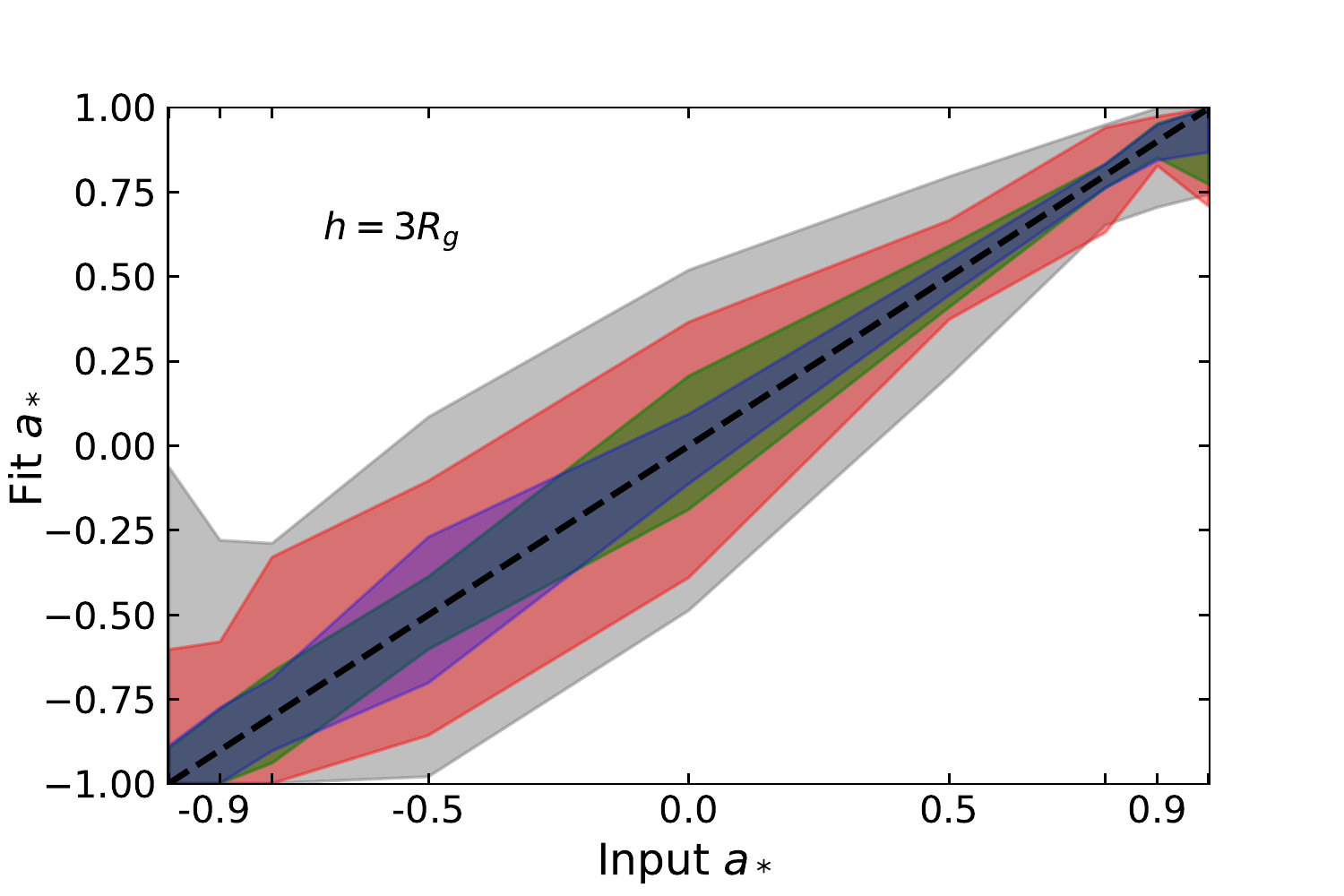}\label{spin1}
		}
		\quad
		\subfigure{
			\includegraphics[width=0.475\textwidth]{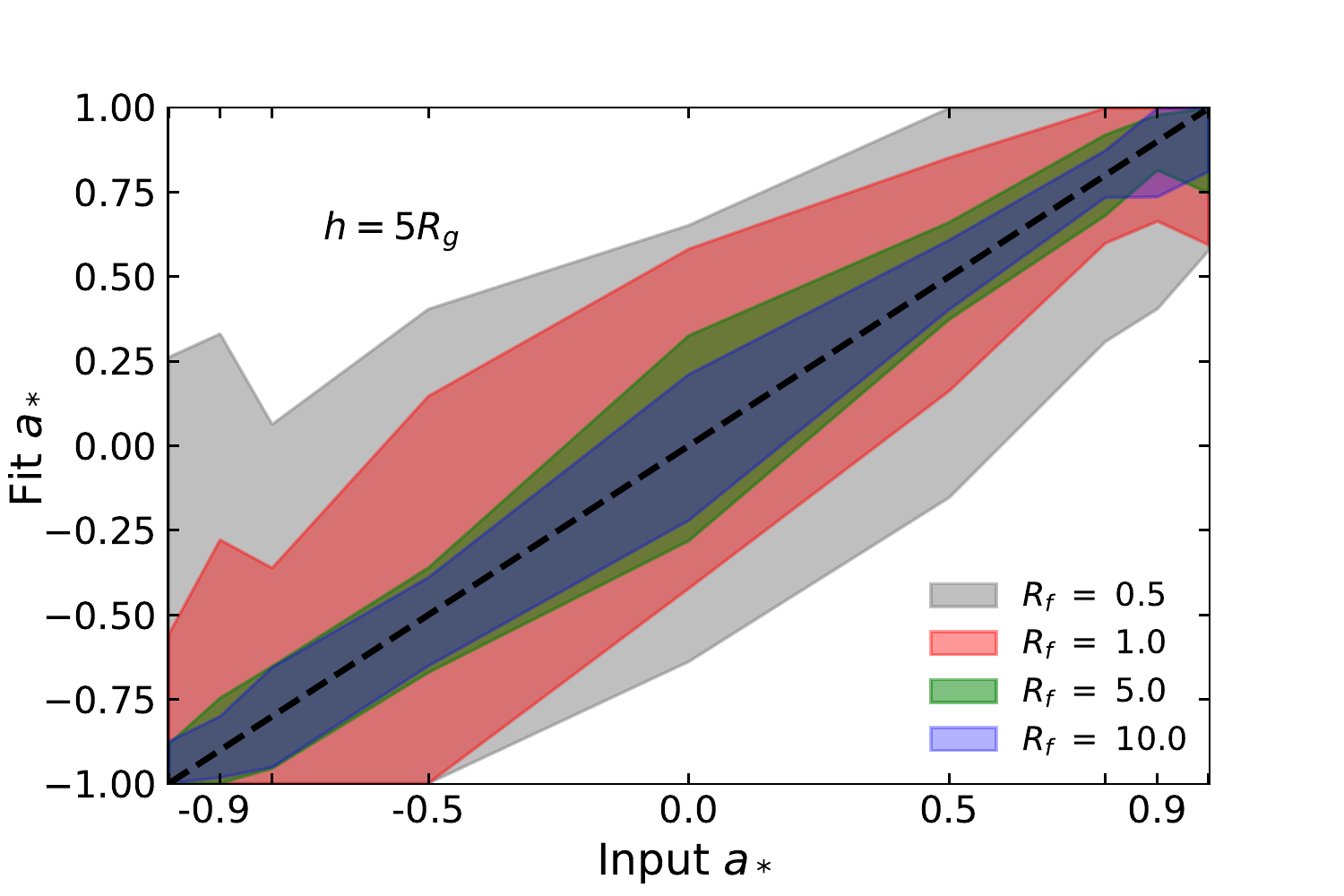}\label{spin2}
		}
		\caption{Spectral fit results for our simulated data measuring spin $a_*$ from all 5,832 simulations of a bright X-ray source, each with an exposure of 100~ks.  The fit parameters are summarized in Table~\ref{params}. Each panel shows results for a different lamppost  height.  The solid-colored intervals depict the 90\% dispersion among the simulations for a given value of input parameter, centered around the mean.  The typical fit statistic for each simulation shown has $\rchi_{\nu}^2
			\leqslant 1.01$. The dashed line depicts the simulation input value for spin.}  
		\label{Spin}
	\end{figure*}
	
	To study the relativistic effects on incoming photons from the
	lamp-post source, we simulate our observations for a bright X-ray corona positioned at 
	each of two lamp heights: $h = 3R_g$ and $h = 5R_g$. We  simulated \textit{NuSTAR} observations of a 
	bright X-ray
	source (at $z = 0$) with a discrete sample space for all parameters of
	interest. 
	Each observation generated was analogous to a 100~ks-long exposure to ensure sufficiently high counts at all energies, and Poisson noise was included in the simulations. After grouping with our adopted method, we were left to fit 301 PHA bins in the desired energy range of 3--79~keV for each spectrum. This number is very small compared to the range of counts we work with here ($\sim 10^6-10^7$). Thus, we overcome the possibility of a fitting bias that using $\rchi^2$-statistics could have imposed as per \cite{cstat2009}. We used nine representative values across the allowed parameter space for spin:
	$\pm$0.998, $\pm$0.9, $\pm$0.8, $\pm$0.5 and 0.0, and used values shown in Table~\ref{params} for the other \texttt{relxilllp} parameters. We explored all combinations in our sampling grid, creating $\sim$5,800 distinct simulations, representing a unique source condition in each. For the parameters $\Gamma$, $log~\xi$ and $A_{\rm Fe}$, the input values have been chosen to represent the physical conditions in most observed systems, as mentioned in \cite{Garcia2013}. While the range for $\Gamma$ in AGN is generally narrower, that for binaries span a wider stretch with very low/hard $\Gamma \sim 1.4-1.5$ to very high/soft $\Gamma \sim 2.5-2.6$ \citep[e.g.,][]{RM2006}. Similarly, AGN disks tend to be less hotter than those in binaries. Not only do spectra having ionization profiles with $\xi \sim 1-20$ appear similar, but also those with $\xi \gtrsim 10^4$ tend to be featureless (closely representing the incident power-law continuum). Iron abundances here are kept at solar and super-solar levels to have prominent {\rm Fe} emission line features, also considering that a sub-solar abundance would not be enough to compensate against the stripping of most of the iron available at higher $\xi$. Over-abundance of iron for explaining X-ray reflection spectra is not uncommon \citep[e.g.,][etc.]{Fabian2006,Rey2012}. The inner disk inclination ($i$) space has been selected to represent systems from being oriented almost face-on to having high relativistic smearing at an almost edge-on view, bearing in mind that the majority of the observed sources can be represented on average between the chosen extremes here.
	
	We let the reflection fraction fit independently
	of the lamppost-prescribed value (i.e., we set \texttt{fixReflFrac
		= 0} in \texttt{relxilllp}).  The fitting procedure was designed such that the
	parameters in each fit initialized at random values that were required to lie beyond $X \pm
	3\sigma_X$, assuming a Gaussian distribution based on the fit covariance around the true value $X$ of each
	parameter, and where $\sigma_X$ was estimated from preliminary fits.
	This approach was adopted to avoid fits becoming stuck in local minima, while at
	the same time allowing the model to explore the parameter space.  We examine this procedure's effect on the overall results in
	Section~\ref{compare}.

	\section{Results}\label{disc}

\begin{figure}
		\centering
		
		\subfigure{
			\includegraphics[width=0.45\textwidth]{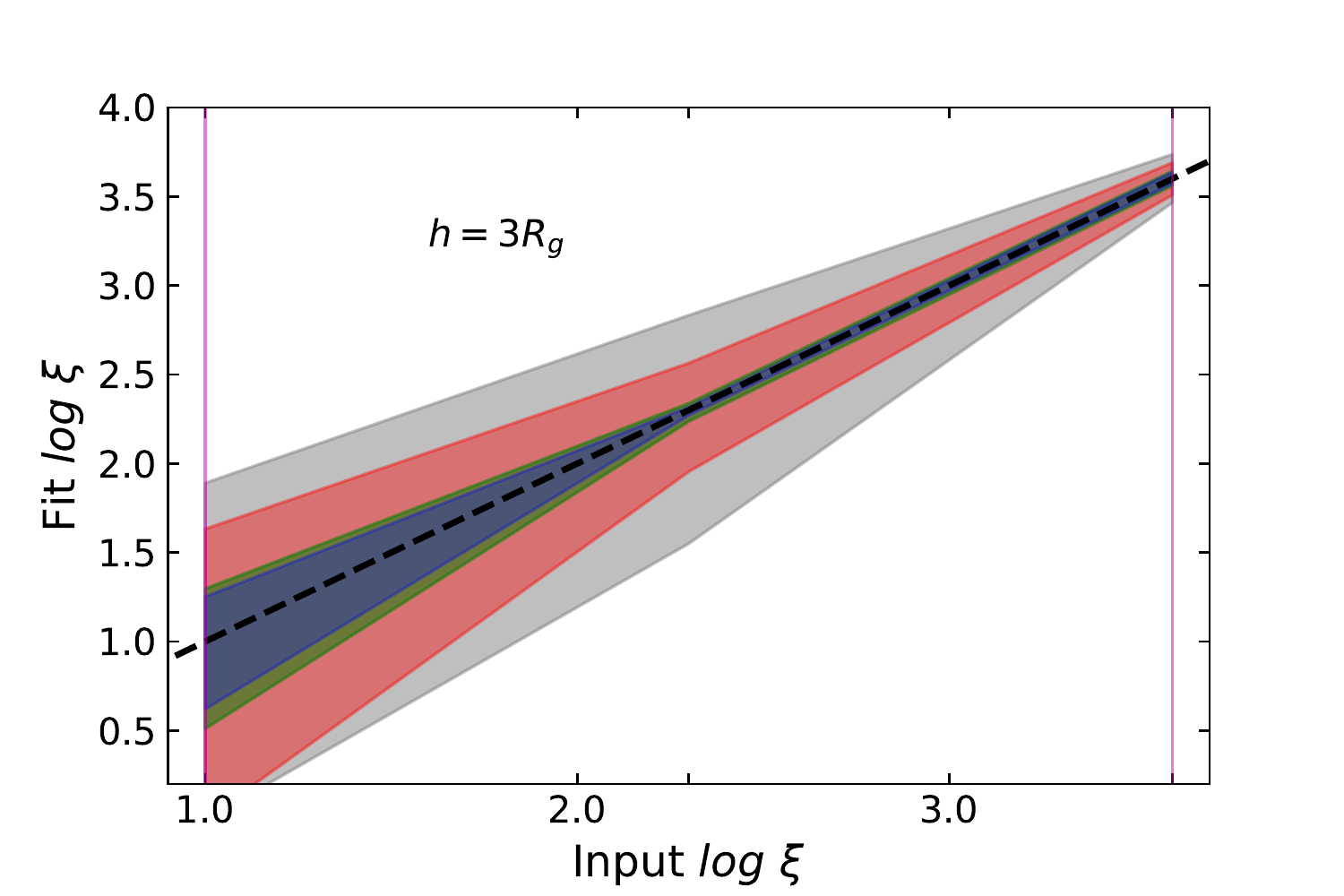}\label{Xi1}
		}
		\quad
		\subfigure{
			\includegraphics[width=0.45\textwidth]{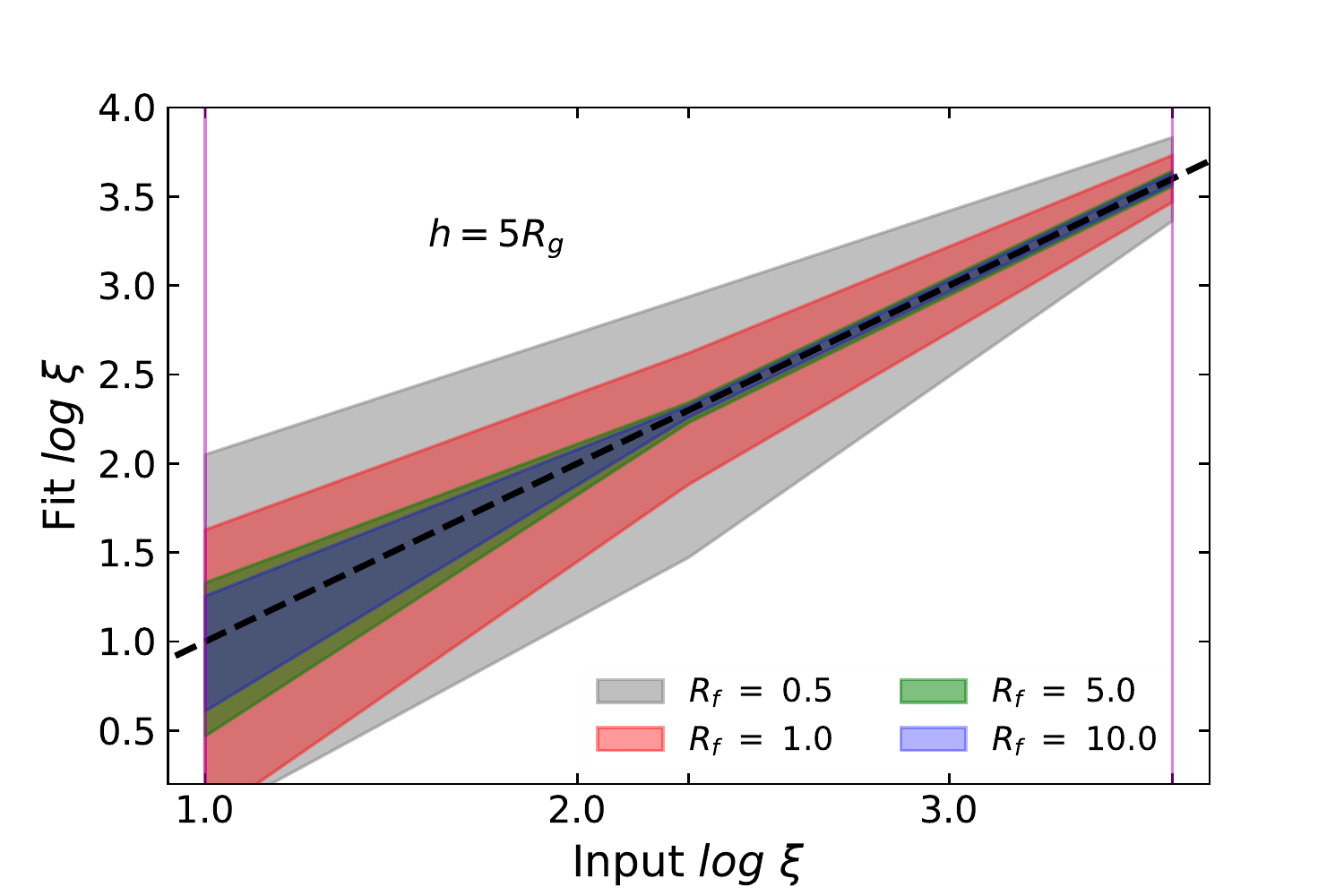}\label{Xi2}
		}
		\caption{Spectral fit results for our simulated data measuring $log~\xi$ from all 5,832 simulations of a bright X-ray source, each with an exposure of 100~ks.  The fit parameters are summarized in Table~\ref{params} with 9 values of $a_*$. Each panel shows results for a different lamppost  height.  The solid-colored intervals depict the 90\% dispersion among the simulations for a given value of input parameter, centered around the mean.  The typical fit statistic for each simulation shown has $\rchi_{\nu}^2
			\leqslant 1.01$. The dashed line depicts the simulation input value for ionization.}\label{Xi}
	\end{figure}

	In Figure~\ref{Spin}, we show the fitted versus inputted values of spin.   Colored regions depict $1.645\sigma$ (90\%) intervals showing the dispersion of the simulation fit dispersion about the average value for spin grid point. All simulations were fitted employing the $\rchi^2$-statistic. We present a parallel set of results showing the fitted values of the ionization
	parameter for the three input values from Table \ref{params} in Figure~\ref{Xi}. The reported $\rchi^2_{\nu}$ is the reduced $\rchi^2$, where $\nu$ stands for the degrees of freedom in the fit.

	Figure~\ref{Spin} shows that the input values are recovered in the fits on the whole.  At a weak reflection signal of $R_f = 0.5$ and a corona located farther from ($h=5R_g$) the compact object, the confidence, for example, on an intermediate spin $a_* = 0.5$ is weaker by a factor of 2 compared to when the corona is at low $h$. Larger reflection fraction
	and larger spin values both lead to tighter parameter constraints.  Tighter constraints are likewise 
	achieved at the lower source height.  This can be understood as each effect separately produces an increase in the amount of relativistic reflection signal.  
	We find that at sufficiently large values of $R_f$ ($\gtrsim 5$), there is negligible change in the confidence interval.  This is because the simulations depict a fixed number of counts, and the proportion of the signal that is conveyed in the reflection component is approximately $\frac{R_f}{R_f+1}$.  For any large value of $R_f$, the reflection signal is essentially the same.  
	
	Similar conclusions can be drawn for the ionization parameter from examination of 
	Figure~\ref{Xi}. For instance, the ionization constraints between $h=5R_g$ and $h=3R_g$ degrade with a ratio of $\sim$1.1--1.7 (progressing from low to high $\xi$, in logarithm scale) at $R_f=0.5$. It is interesting that higher dispersion seems to occur at low ionization ($log~\xi \lesssim  2$). This is explained by \cite{Garcia2013}, where it can be seen that the difference between reflection spectra in the \textit{NuSTAR} band are quite minor at such low values of ionization.  By contrast, at high ionization, the spectra are more readily distinguished, as reflected in the fits. It is, however, harder to recover a hotter disk at higher source height at weak reflection signal (wider relative confidence region for hotter disk) compared to when the primary source is closer because of lesser reflection signal received from the inner disk as the point source moves higher up and away. In general, for the model parameter the
	dispersion intervals here, again, are slightly narrower for lower $h$, and they reduce appreciably 
	with growing $log~\xi$ or increasing $R_f$.  

	\section{Discussion} \label{compare}
	
	\begin{figure} \label{spinBG}
		\centering
		
		\includegraphics[width=0.475\textwidth]{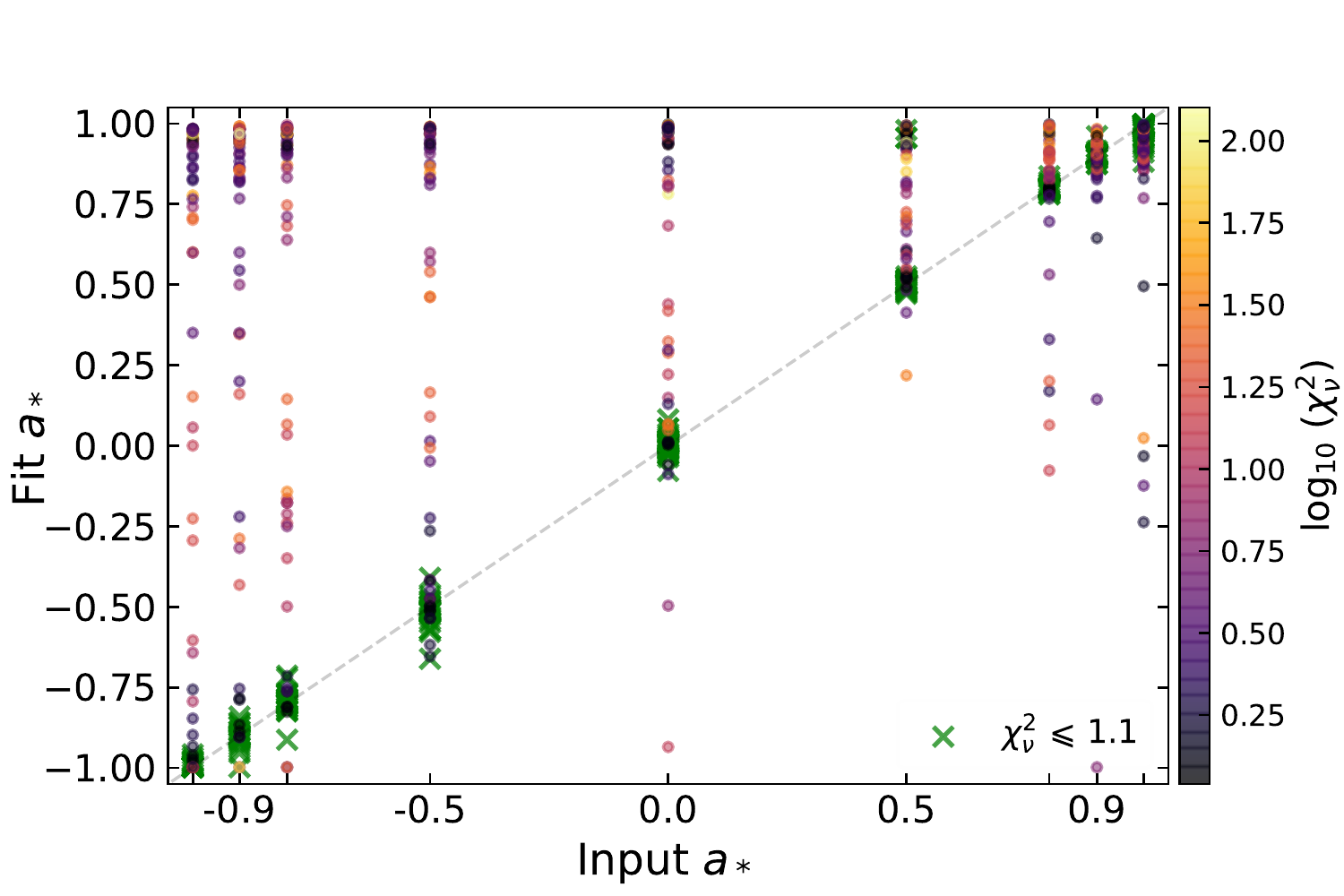}
		
		\caption{Spin fitting results following the approach of BG16 at our total counts.  We show the fitted spin ($a_*$) versus its input for model settings of $R_f=5.0$ and $h=3R_g$. Each data point is color-coded according to the logarithm of its reduced $\rchi^2$. Good fits with $\rchi^2_{\nu} \leqslant 1.1$ have been shown separately to compare with BG16 work.} \label{spinBG}
	\end{figure}

	\begin{figure*}[!t]
		\centering
		
		\includegraphics[width=1.0\textwidth]{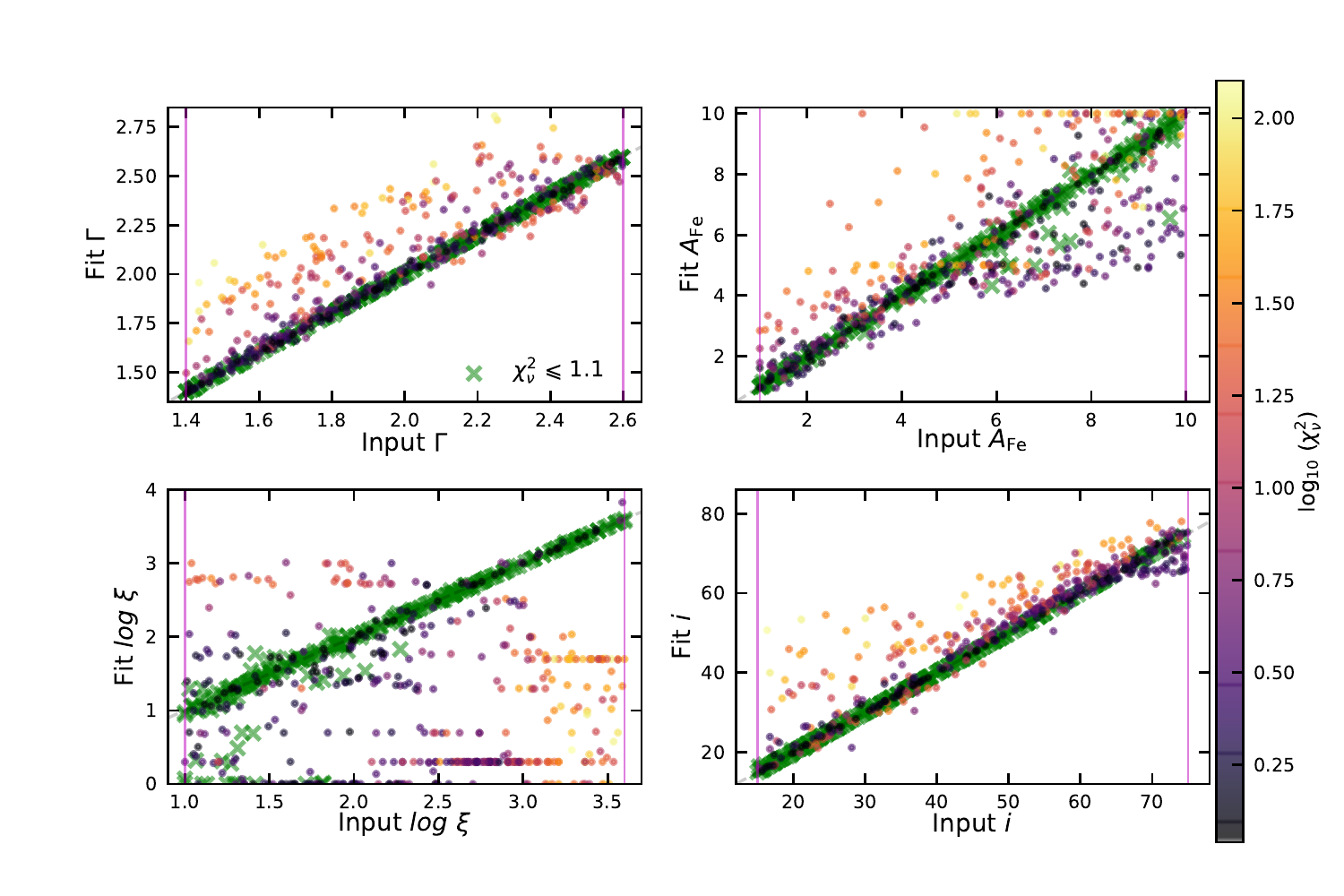}\label{spin2}
		\caption{Results using the approach of BG16, as in Fig.~\ref{spinBG}, but showing $\Gamma$, $A_{\rm Fe}$, $log~\xi$, and $i$.  
			Although the color-coding indicates overestimation in $\Gamma$ and $i$ and underestimation of $log~\xi$, we find that none of the good fits of these parameters, other than a few $log~\xi_{inp}=2.0$ underestimations, are under/overestimated at our signal and with our grouping approach.}\label{otherBG}
	\end{figure*}
	
	A similar investigation to ours was recently presented by \cite{BG} (BG16 hereafter).  BG16 analyzed over 4,000 simulated spectra of AGN using {\sc relxill}, with $3.5 \times 10^5 \pm 10^3$ net counts each between 2.5--10~keV with \textit{XMM Newton} \cite[EPIC-pn;][]{Struder2001} and 10--70~keV with \textit{NuSTAR}. They determined the effects on estimates of relativistic reflection parameters with Fe-line fitting in AGN X-ray astronomy and concluded that measuring such parameters accurately can be of great challenge, especially spin ($a_*$) parameter which they found can be decisively recovered for $a_*>0.8$ with an accuracy of $\pm$0.1.

	Like us, BG16 explored {\sc relxill} behavior over a wide range of parameter space by 
	simulating high-signal observations. However, unlike us, they initialized all their model fits at 
	a single starting set of conditions (which could be arbitrarily far from or really close to the inputs).  This type of fitting could be interpreted as what a particular user would do when there is no information regarding the parameters that characterize the system. On the contrary, our approach tries to simulate the other extreme possibility, in which the user has a good rough estimate of the parameters, based on previous analyses or measurements performed with other techniques. They report to have performed error checks on all fit parameters for one set as a verification step in the 2.5--70~keV band and that there were no significant differences found from their published results that were obtained only by running the fit. The effect of doing so without 
	enforcing rigorous exploration of parameter space (e.g., via {\sc error}, {\sc steppar}, or MCMC approaches) can lead to erroneous results because
	the fit procedure can easily become trapped in a local minimum, which BG16 mention as well. Indeed, this shortcoming is not unique to BG16's approach, but is a caution that should be marked by observers to ensure that parameter space is explored   
	intensively and sufficiently. In Section \ref{degen}, we also used small step sizes (equal to or smaller than the predetermined $\sigma_X$) for the parameters of interest so that the fit can explore the parameter spaces better, although leading to the code running comparatively slower. We have attempted to elucidate the differences, with lamppost geometry, between results employing our fitting approach and those from replicating BG16 procedure as a representation of what is being done in day-to-day X-ray data analyses.

	We chose to use our well-recovered set with $R_f=5.0$ and
	$h=3R_g$ for reference, and adopt all 9 of our values of spin parameter.   We simulated and fitted 
	the same number of observations as in Section~\ref{degen} for a single $R_f$ and $h$ combination with
	our source counts. The primary difference between this examination and BG16's is 
	that we kept the input spin values discretized, whereas they selected values at random (but from the same range we adopt).
	When fitting, all parameters were initialized according to the BG16 approach (i.e., their ``Test A'' set: $a_*=0.5$, $i=30$~deg, $\Gamma=2.0$, $\xi=75$
	erg~cm~s$^{-1}$, and $A_{\rm Fe}=3$).  
	We again binned the simulated data to 3-times oversample the detector energy resolution. We ran errors on all fit parameters.

	Figure~\ref{spinBG} shows the results for recovering black hole spin with the representational fitting approach above.  We show the $\rchi^2_{\nu} \leqslant 1.1$ good fits separate from the color-coded scatter. We can clearly see here that only a minority ($\approx$~35\%) are actually good fits in the high signal sample considered that we further inspected are not stuck in local minimum beyond a representational 10$\sigma$ statistical significance over all fit parameters. Here, error checks push to span the parameter space better and are able to converge to good best-fit statistics. However, evidently, there is a much larger scatter of fits with $\rchi^2_{\nu} > 1.1$ and 
	many spurious results appear to muddy the determination of $a_*$.  We find that it is much more likely for the  
	fit to miss entirely and when it does to overestimate the spin. These are the larger share of bad fits stuck in local minima, and can be compared for a contrast with results from our fitting approach with similar parameter combination in Figure~\ref{Spin}. Figure~\ref{otherBG} shows
	the scatter in the other parameters, and we can see that both $\Gamma$ and $log~\xi$ have large scatter similar to that shown by BG16. The good fits show no under/overestimation in the photon index, the iron abundance and the inner disk inclination parameters. We can also see a few good fits with input $log~\xi < 2.0$  being underestimated, but it can be explained because of the spectral similarity at low ionization mentioned in Section~\ref{disc}.
	
	As apparent in Figure~\ref{otherBG}, and confirmed by exploration of the data, the primary problem in question with the fits in local minima arises due to incorrect measurements of $log~\xi$ pushing
	spin towards poor estimates.  The nonlinear behavior exhibited by ionization can well be a problem in effectively analyzing data without a sound fitting procedure. To test this hypothesis we performed a qualitative
	examination on all four parameters shown in Figure~\ref{otherBG}.  We excluded all fits $\leqslant |2\sigma|$ confidence around an ideal recovery (i.e., fit
	= input) for each input value for $a_*$, and mapped the remainder (outliers all) to the Figure~\ref{otherBG} fit parameters.  The results confirmed that the worst spin
	fits were associated with very large mismatches in the
	ionization parameter.  Iron abundance, followed by $\Gamma$ 
	and $i$, were minimally associated with causing outliers in spin.  
  This is likely because the spectral changes associated with evolving ionization are not necessarily smooth and continuous, unlike e.g., changes associated with $i$ or $A_{\rm Fe}$ (see e.g., \citealt{Garcia2013}).   

	\subsection{Line and Continuum Features Trading Off} \label{gamma-xi}
	
	\begin{figure*}[!t]
		\centering
		
		\subfigure{
			\includegraphics[width=0.45\textwidth]{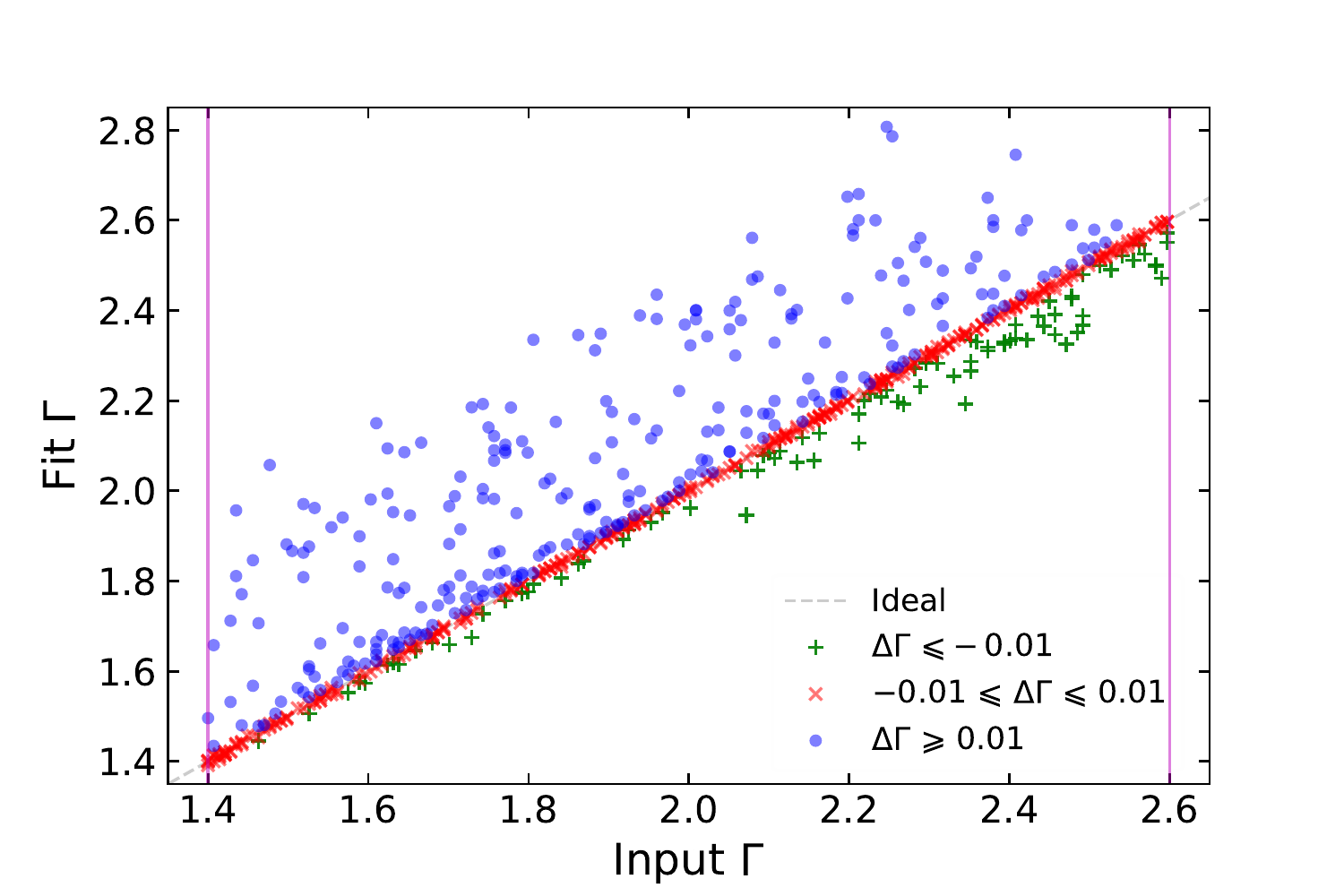}
		}
		\quad
		\subfigure{
			\includegraphics[width=0.45\textwidth]{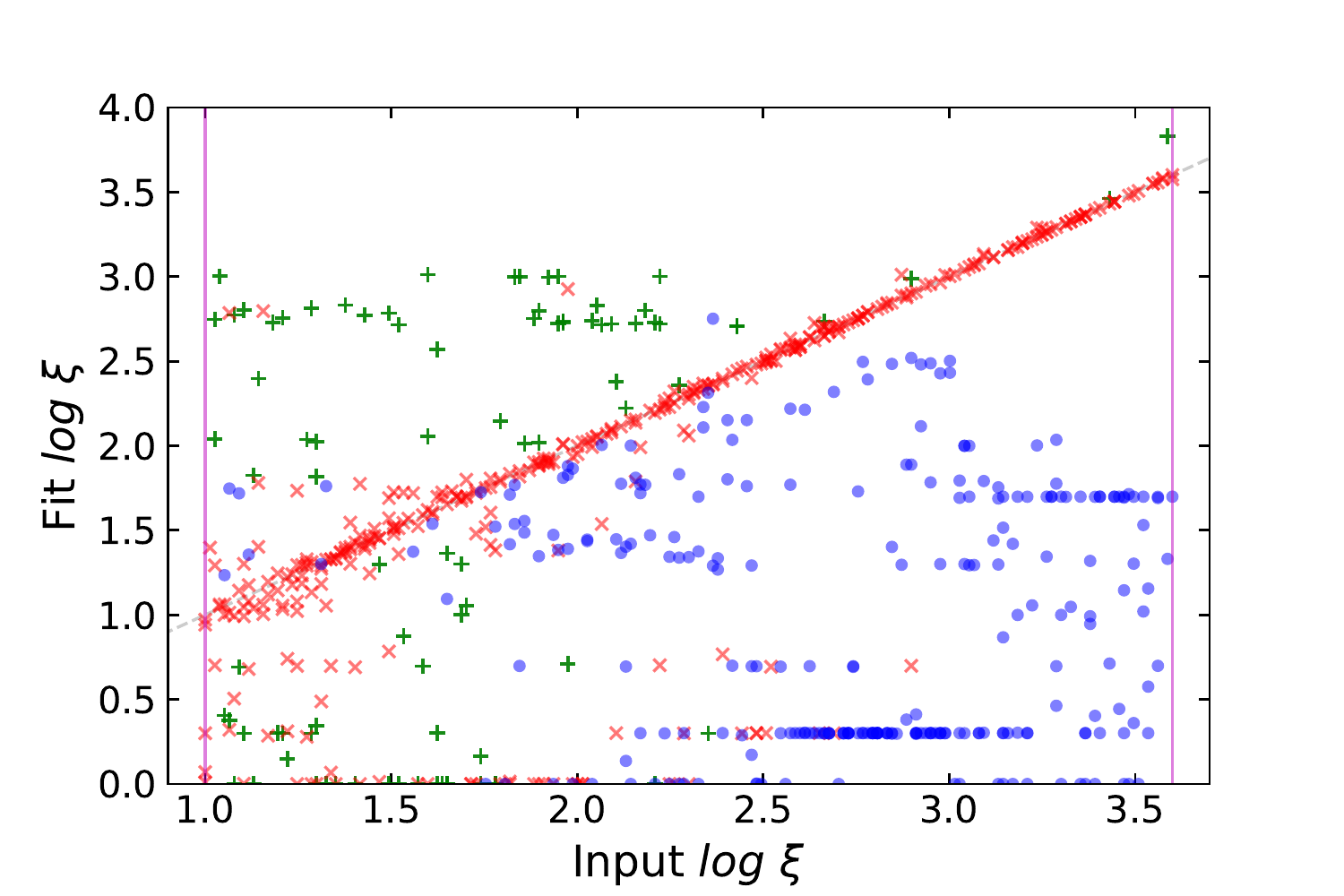}
		}
		\caption{The left upper and lower panels of Fig.~\ref{otherBG} color-coded 
			based on the difference $\Delta \Gamma \equiv \Gamma_{\rm fit} - \Gamma_{\rm input}$.  There is clear correspondence between the differences in $\Gamma$, and in $log~\xi$.} \label{delGamma}
	\end{figure*}
	
	Having noticed that $\Gamma$'s distribution of spurious fits in Figure~\ref{otherBG} is strongly skewed to fitting larger values of $\Gamma$, and revealing surprisingly large discrepancies up to $\gtrsim 0.1$ (far larger than statistical errors), we explore the relationship between $\Gamma$ and the nonlinear parameter $log~\xi$ from our results with the representational fitting approach.      
	
   To examine this behavior, in Figure~\ref{delGamma} we have color-coded the relationships of $log~\xi$ and $\Gamma$ from Figure~\ref{otherBG}. Blue depicts cases with $\Delta\Gamma > 0.01$, green shows $\Delta\Gamma < -0.01$, and red gives the remaining cases (all good fits) in which the fit is close to the input value. Note that fits which greatly underestimate $log~\xi$ correspond to fits which overestimate $\Gamma$, and vice versa.  
	
	This correlation can be understood again from the reflection principles outlined in 
	\cite{Garcia2013}.  Low values of $log~\xi$ emit less at soft X-rays and so produce harder X-ray spectra.   In order for the fit to compensate for the harder signal in the reflection, $\Gamma$ is increased.   In this way, there is correlated trade-off between the parameters describing narrow reflection features (i.e., $log~\xi$), and the X-ray continuum ($\Gamma$).
	
	In order to see how the results fare at lower counts, we reproduced the dispersion shown in figures~\ref{Spin} and \ref{Xi} for the set with $R_f = 5.0$ and $h = 3R_g$, the scatter in figures~\ref{spinBG} and \ref{otherBG}, and the mapping in Fig~\ref{delGamma}-- all at BG16's total counts. As expected, the dispersion in spin and ionization increased at all input values of $a_*$ and $log~\xi$ with our fitting approach, although not deviating anywhere from the input line but having noticeably wider confidence at lower input values. The trend in the scatter plots with the representational fitting approach at lower counts was similar to those shown in this work at higher counts. The percentage of good fits not stuck in local minima increased (to $\approx$~48\%) but had much larger deviations from the input lines. We observed that most of the good fits sticking around the input line in this case could be attributed to well-recovered $i$ and $\Gamma$, while $log~\xi$ again remained the primary culprit in creating spurious results with higher number of good $\xi$ fits being underestimated. This ascertains the need to adopt a proper fitting approach when testing model parameters at any range of total counts.

	\subsection{Bias, Binning, and Statistical Methods}\label{StatTest}
	
	\begin{figure*}[!t]
		\centering
		
		\subfigure{
			\includegraphics[width=0.475\textwidth]{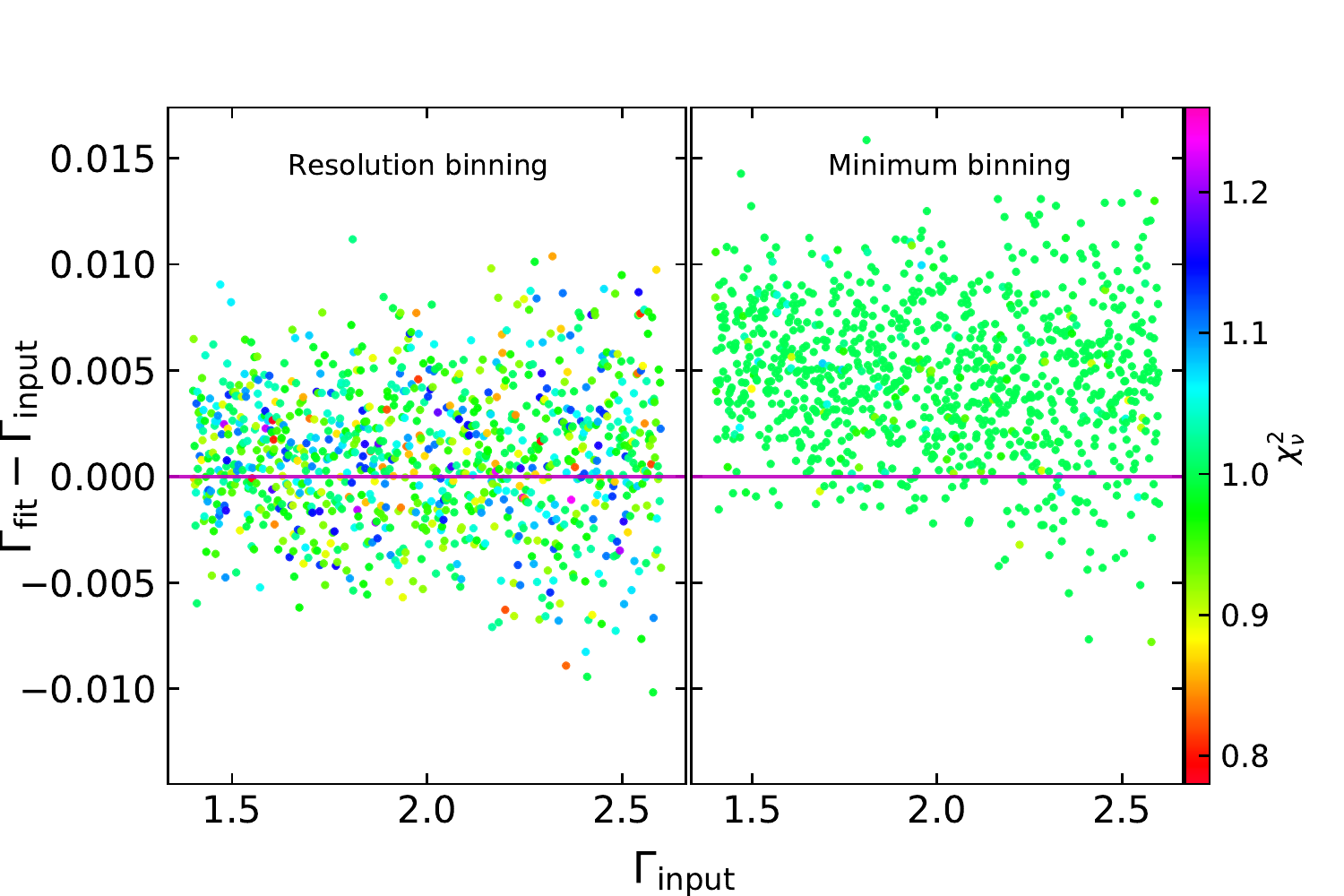}
		}
		\quad
		\subfigure{
			\includegraphics[width=0.475\textwidth]{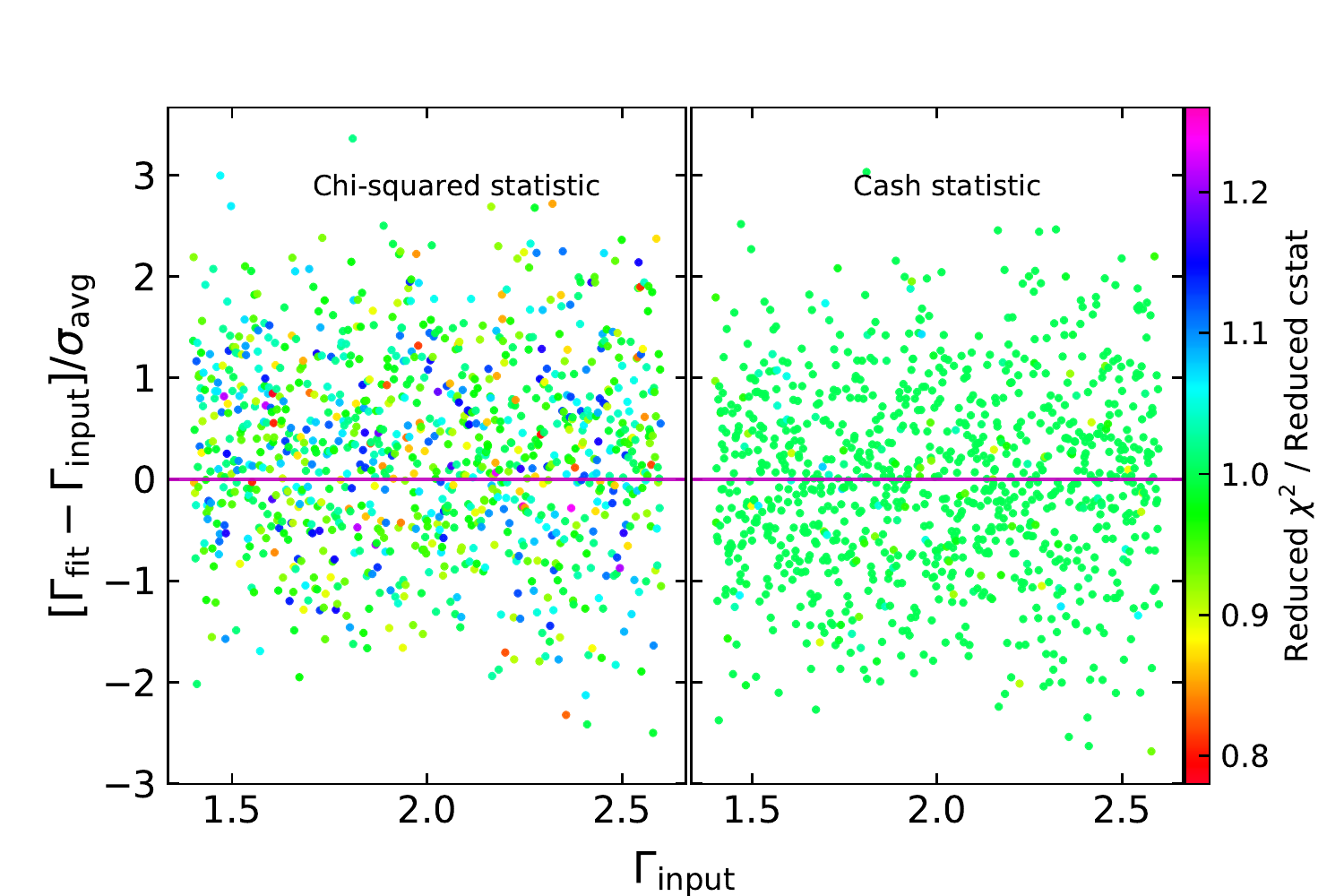}
		}
		
		\caption{Color-coded fits from 1,000 Seyfert 1 spectra simulated with \texttt{POWERLAW} model with \textit{NuSTAR}.\\
			\textit{Left}: Scatter plot for the difference $\Gamma_{fit} - \Gamma_{input} (= \Delta \Gamma)$ for two sets, one with data oversampled according to our resolution-binning method (left pane), and the other with a minimum grouping of 25~cts~bin$^{-1}$ (right pane).\\ \textit{Right}: Scatter plot for the difference $\Delta \Gamma$ in terms of $\sigma$ for the same set of resolution-binned data, fit one time each using $\rchi^2$ statistic and \texttt{cstat}. $\sigma_{avg}$ refers to the average statistical error bar for each point in each scatter plot. The scatter $>|2\sigma|$ is negligible. The color-bar represents $\rchi^{2}_{\nu}$ and reduced-\texttt{cstat}.} \label{fig6}
	\end{figure*} 
	
	\begin{figure*}
		\centering
		
		\subfigure{
			\includegraphics[width=0.475\textwidth]{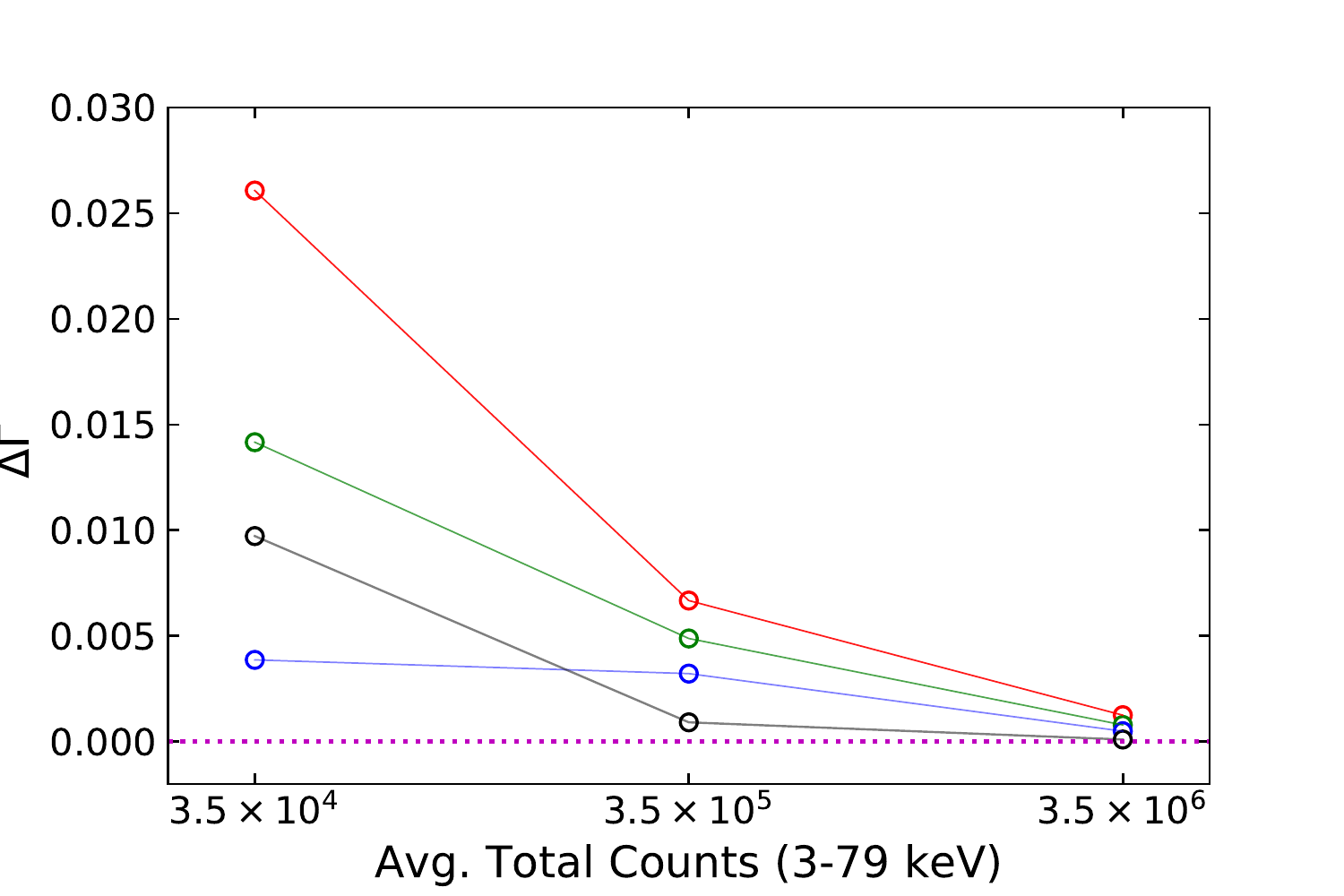}
		}
		\quad
		\subfigure{
			
			\includegraphics[width=0.475\textwidth]{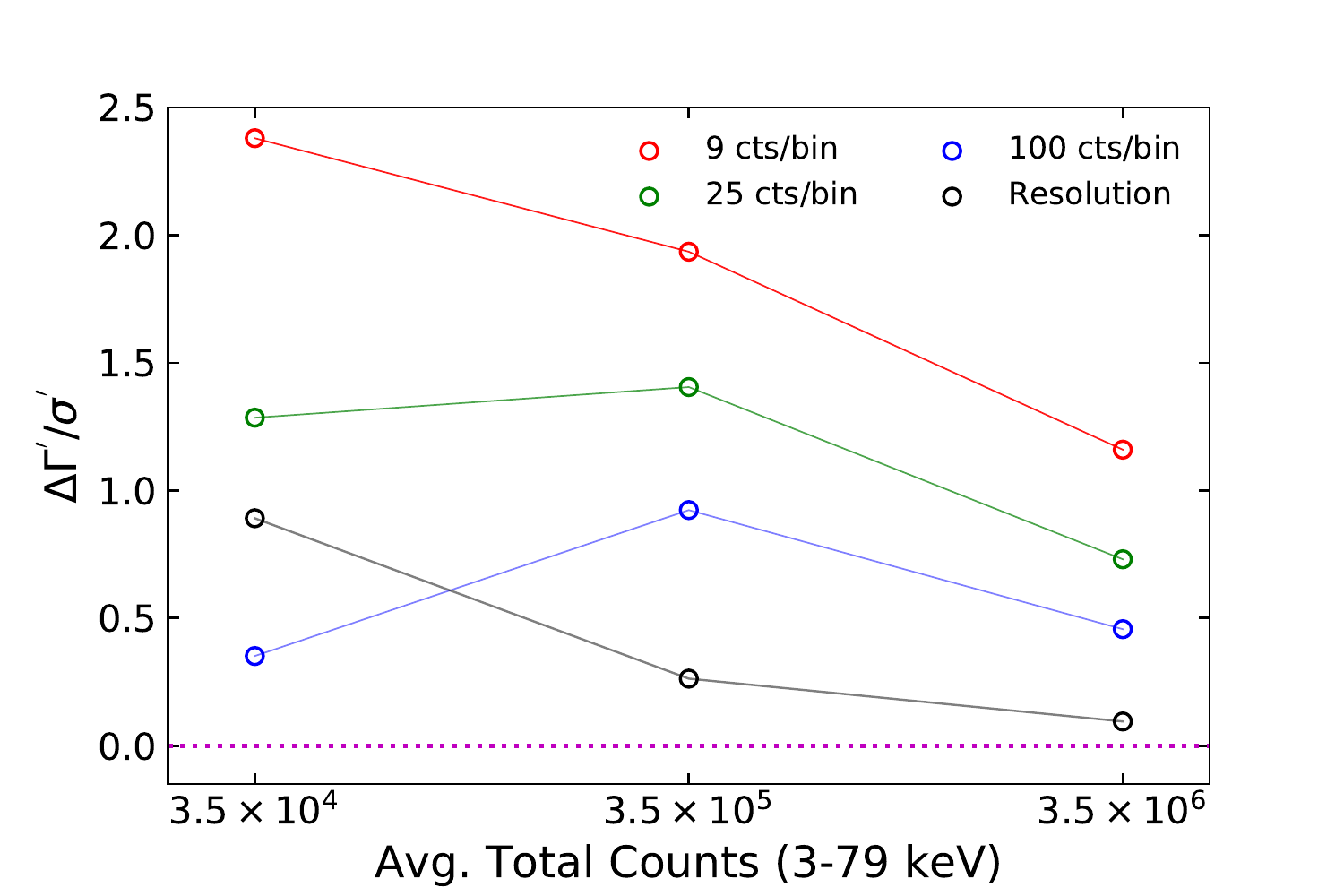}
		}
		
		\caption{Results from pure power law simulations illustrating the effect of varying the total signal and the grouping.  Each point is obtained from a $\rchi^2$-minimizing fit, and the results are compared to a reference analysis using the Cash statistic. The left panel shows the absolute offset in $\Gamma$ ($\Delta\Gamma' \equiv
			\Gamma_{\rm (fit,\rchi^2)} - \Gamma_{\rm (fit,cstat)}$), and the right panel depicts the bias in terms of its statistical significance.} \label{fig7}
	\end{figure*}
	As seen from Section~\ref{gamma-xi}, $\Gamma$ seems to be overestimated in the case of assessing blurred reflection using {\sc relxill} over a broad range of counts with the representational fitting approach.  While we described above the way in which an offset in $log~\xi$ can introduce a correlated shift in the continuum $\Gamma$, this effect (also reported by BG16 for $\Gamma$ at $\rchi^2_{\nu} \leqslant 1.1$) is more subtle and is rooted in the data treatment because in our case it shows up, for both range of counts, only at $\rchi^2_{\nu} > 1.1$. In this section, we illustrate how the use of $\rchi^2$ rather than Poisson statistics can fully account for this reported bias. 
	
	To demonstrate the problem, we have produced 1,000 simulations using a pure power law model for the same \textit{NuSTAR} FPMA response to look into the photon index independently, with randomly sampled values of $\Gamma$ ranging between 1.4--2.6.  Each spectrum contains 3.5$\times10^5$ counts (within Poisson limits) across the useful 3--79~keV band. 

	These simple simulations were then binned using two approaches:  (1) the data were binned according to our default procedure, oversampling the detector resolution by a factor of 3, or (2) the data were grouped to achieve a minimum of 25 counts per bin, and the detector resolution was not taken into consideration.  We also test the importance of the Gaussian approximation inherent in $\rchi^2$ fitting, in both cases, by comparing results achieved by $\rchi^2$ minimization with parallel results achieved using the proper Cash statistic (\citealt{Cash1979}; termed ``cstat'' in {\sc xspec}), which is appropriate for Poisson-distributed data.  
	
	Figure~\ref{fig6} illustrates our results. The leftmost pair of panels depicts two sets of $\rchi^2$ fits using the two binning options, presenting the difference between fitted and inputted $\Gamma$.  A strong bias in $\Gamma$ is clearly evident in the ``minimum binning'' panel, which relates to the aforementioned binning option (2).  In this instance, one infers that $\Gamma$ is generally too high, by a factor $\sim 0.005$.   The precise value of this offset turns out to be a function of the binning and also of the total signal in the spectrum, with the bias becoming worse when the signal is sparser and the counts per bin fewer.  As it can be seen, nearly all fits in the minimum binning case have good converging statistics. However, the majority appear to be stuck in local minima, largely showing overestimate in $\Gamma$ owing to the sampling performed. Importantly, our adopted resolution binning appears comparatively immune to this bias and is clearly preferable.
	
	However, the rightmost pair of panels show that when taking a more careful look, as revealed here by scaling the difference in $\Gamma$ by the statistical error, even the resolution-binning case suffers from a small bias when employing the approximate (but most widely-used) $\rchi^2$-statistic compared to the furthest right panel which shows the same using the Cash statistic.  Note the Cash-statistic distribution is centered on zero with good converging statistics, while the $\rchi^2$ case is very slightly offset to larger fitted $\Gamma$ (a minor shift of order $\sim 0.2 \sigma$).  Given the very minor scale of the bias associated with our adopted approach, our earlier results from Section~\ref{disc} are quite acceptable.

	We further illustrate the dependence of the bias in $\Gamma$ on the total counts and degree of binning in Figure~\ref{fig7}.  Here, we select a single representative value of $\Gamma=2.0$ as a fiducial value and have varied the counts per spectrum by an order of magnitude in either direction in a smaller set of simulations that depict different data groupings, where the minimum number of counts per bin is 9, 25, or 100.  Using a fit with the Cash statistic as our reference point, we show the resulting bias in $\Gamma$ in the left panels and compare that to its significance in the right panels.  We present the results for our adopted resolution-binning as well.

	This minimum binning approach is commonly adopted for the analyses of observational X-ray data in order to have ample counts in each data bin, and overlooks the detector resolution. The overestimation found in $\Gamma$ can be evidently seen due to this. Apart from insufficient binning of data, using Gaussian-assuming $\rchi^2$-statistic to model Poisson-distributed data could also be a cause. Accordingly, we recommend using the Cash statistic when possible, and suggest further that it is advisable to bin the data according to the detector resolution (refer to page \pageref{humphrey} for references on the choice of statistics and data binning).

	\section{Conclusions} \label{conc}
	
	We have simulated \textit{NuSTAR} data of a bright X-ray source to determine how well  \textsc{relxill} can practically determine spin and ionization parameters.  We have adopted the simplest case of lamppost geometry.  Our results are summarized below:
	
We find that all model parameters are well-recovered in the fits to our reflection simulations, and that the precision is improved at higher $R_f$, higher spin and lower $h$, because each increases the signal-to-noise in relativistic reflection features. Recovering retrograde spin is more difficult compared to their prograde counterparts owing to the lower reflection signal received because of the position of the inner disk radius $R_{\rm in}$, which we have fixed at the innermost stable circular orbit (ISCO). $R_{\rm ISCO}$ for a counter-rotating black hole tends to increase outwards with more retrograde values, thereby limiting reflection from the inner regions of the disk. Further complications may arise because of lowered relativistic bending at higher source heights.

		We carried out an examination of adopting fitting methods while analyzing X-ray data based on results from a similar work done by \cite{BG}, across our selected $10^6-10^7$ range of total counts for testing the estimates in {\it NuSTAR}'s effective energy range. BG16 show that the model {\sc relxill} yields poor estimates for the relativistic and disk parameters when a representational fitting approach like they adopt is employed. We also performed the examination at lower ($\sim 10^5$) counts and, to our expectations, obtained similar results. We find that the intrinsic bias imposed by such a fitting approach, extended to any range of detected counts, can be minimized by instead adopting to fit data with some predetermined estimates around the true parameter configuration, owing to the complexity of the model itself. This points to the efficiency of the fitting method we adopt in Section~\ref{degen}, and also stands to support the concern expressed by BG16 over such common X-ray fitting methods. Furthermore, parameters like $log~\xi$, which has a nonlinear impact on spectra, require thorough stepping through their parameter spaces in order to avoid internal subtle adjustments among parameters that can easily make the user believe in misleading good fits.
		
       On top of the fitting method chosen, picking the data binning scheme also seems to impact results when using the Gaussian-assuming $\rchi^2$-statistic, which can be seen from the affect on the global parameter $\Gamma$ in Section~\ref{StatTest}. In the case of data being Poisson-distributed, it is crucial to employ a sound binning methodology. Then again, using Cash statistic is preferably the better choice as it is suitable for treating Poissonian data. We therefore advise that data should be binned according to the detector resolution and the Cash statistic employed when possible (ensuring at least 1 count per bin at very faint detections when the noise is purely Poissonian).

		The results put forth in this work have been determined in the most ideal case: using one module and the broad bandpass of {\it NuSTAR} with Poissonian background, devoid of effects like inter-stellar medium (ISM) absorption $<2$~keV, to assess blurred reflection data using lamppost irradiation. While the results depict the efficiency of {\sc relxill} in being able to constrain spin and ionization very well, the conditions can be conservative not only because of the geometry adopted but also in the added sense that there is no unique ``correct'' method when adopting a fitting approach in order to probe a model's efficacy. In addition to the selection of a sound fitting methodology, the extent to which the user can avoid local minima relies on the fitting algorithm, data sampling and the degrees of freedom in the fit that may complicate the situation altogether. The user, however, needs to be cautious since results obtained may in fact shadow the true nature of parameter constraints and degeneracies (which we have not probed into for the current work) involved intrinsically, a common problem while analyzing real data.
		
		Simultaneous fitting with simulations from instruments operating at lower energies, like {\it XMM Newton} or {\it Suzaku}, can significantly improve our constraints, and may serve as an extension to the current work. Similar work in \cite{Garcia2015a} showed that the constraints in $E_{\rm cut}$ in {\sc relxill} improved with the use of a broader waveband encompassing soft energies. An alternative in the case of data analyzed from observations could be ``pgstat'' which reads Gaussian background with Poissonian detection (see Appendix B in {\sc xspec} manual). The implementation effect of this fit statistic can be tested with real data or a simulated Gaussian noise since background can definitely be not just Poisson-distributed in reality. But this is beyond the goals and scope of this work.
		
	\acknowledgments{We thank the anonymous referee for the extensive suggestions in improving the paper. We also thank Sourabh Nampalliwar for helping with PyXspec issues on the clusters at the Department of Physics, Fudan University. The work of K.C. and C.B. was supported by the National Natural Science Foundation of China (NSFC, Grant No.~U1531117) and Fudan University (Grant No.~IDH1512060). K.C. also acknowledges the support from the Chinese Scholarship Council (CSC), Grant No.~2015GXYD34. J.A.G. acknowledges the support from NASA, Grant No. NNX15AV31G. J.F.S. was supported by Einstein Fellowship Grant PF5-160144. J.A.G. and C.B. also acknowledge the support from the Alexander von Humboldt Foundation.}

	\bibliography{RELXILLpaperK}

\end{document}